\documentclass{article}

\usepackage{arxiv}
\usepackage{amsmath}
\usepackage[utf8]{inputenc} 
\usepackage[T1]{fontenc}    
\usepackage{hyperref}       
\usepackage{url}            
\usepackage{booktabs}       
\usepackage{amsfonts}       
\usepackage{nicefrac}       
\usepackage{microtype}      
\usepackage{lipsum}
\usepackage{graphicx}

\usepackage{epstopdf}
\usepackage{pdflscape}
\usepackage{graphicx}
\usepackage{tabularx}
\usepackage{algorithm}
\usepackage{algorithm}
\usepackage{algpseudocode}
\usepackage{array}
\usepackage{verbatim}
\usepackage{algorithmicx}
\usepackage{subfig}
\usepackage[autostyle]{csquotes}
\usepackage{algpseudocode}

\usepackage{authblk}

\title{Real-Time Lightweight Chaotic Encryption for 5G IoT Enabled Lip-Reading Driven Secure Hearing-Aid}

\author[1]{Ahsan Adeel*}
\author[2]{Jawad Ahmad}
\author[1]{Amir Hussain}
\affil[1]{Department of Computing Science and Mathematics, University of Stirling, FK9 4LA, UK}
\affil[2]{School of Engineering and Built Environment, Glasgow Caledonian University, Glasgow, G4 0BA, UK  *ahsan.adeel@stir.ac.uk }

\begin{document}
\maketitle

\begin{abstract}
Hearing-loss is the third most common chronic health condition. More than 10 million people in the United Kingdom (UK) and approximately 360 million in the world are suffering from a debilitating hearing loss, with the number estimated to rise to 14.5m by 2031. Existing audio-only hearing-aids are known to perform poorly in noisy situations where overwhelming noise is present. Next-generation audio-visual (lip-reading driven) hearing-aids stand as a major enabler to realise more intelligible audio. However, high data rate, low latency, low computational complexity, and privacy are some of the major bottlenecks to the successful deployment of such advanced hearing-aids. To address these challenges, we envision an integration of 5G Cloud-Radio Access Network (C-RAN), Internet of Things (IoT), and strong privacy algorithms to fully benefit from the possibilities these technologies have to offer. The envisioned 5G IoT enabled secure audio-visual (AV) hearing-aid transmits the encrypted compressed AV information and receives encrypted enhanced reconstructed speech in real-time which fully addresses cybersecurity attacks such as location privacy and eavesdropping. For security implementation, a real-time lightweight AV encryption is utilized. For speech enhancement, the received AV information in the cloud is used to filter noisy audio using both deep learning and analytical acoustic modelling (filtering based approach). To offload the computational complexity and real-time optimization issues, the framework runs deep learning and big data optimization processes in the background on the cloud. Specifically, in this work, three key contributions are reported: (1) 5G IoT enabled secure audio-visual hearing-aid framework that aims to achieve a round-trip latency up to 5ms with 100 Mbps datarate (2) Real-time  lightweight  audio-visual encryption based on piece-wise  linear  chaotic  map (PWLCM),  chebyshev  map,  secure  hash, and  a  novel substitution  box  (S-Box) algorithms (3) Lip-reading driven deep learning approach for speech enhancement in the cloud. The effectiveness and security of the proposed secure AV hearing-aid is extensively evaluated using widely known security metrics such as correlation coefficient,  entropy,  contrast,  energy, number of pixel change rate (NPCR) and unified average change intensity (UACI). Lip-reading driven deep learning approach for speech enhancement is evaluated under four different dynamic real-world commercially-motivated scenarios (cafe, street junction, public transport, and  pedestrian area) using benchmark Grid and ChiME3 corpora. The critical analysis in terms of both speech enhancement and AV encryption demonstrate the potential of the envisioned technology in acquiring high quality speech reconstruction and secure mobile AV hearing aid communication. 
\end{abstract}

\section{Introduction}
Hearing impairment is a hidden disability with no painful symptoms. People with serious hearing-loss find themselves socially isolated and depressed with more negative consequences including headaches, muscle tension, increased stress, insecurity, and sadness. Hearing aids are the most widely used devices for the majority of hearing losses compensation. The global hearing aid industry, estimated around US \$6 billion, is expected to grow at 6 percent annually through 2020, according to the market research firm MarketsandMarkets. However, existing hearing aids often work poorly for speech in noise. Current devices make the signal more audible but remains deficient to restore intelligibility  i.e., no improvement in SNR. Thus, the existing audio-only hearing-aids are not robust to reverberation and intelligibility wins at the cost of higher cognitive load in noisy environment \cite{ruggles2012ignitability}\cite{kortlang2016combination}. 

Despite decades of research, only few speech enhancement algorithms could reliably increase the intelligibility of speech in noise, especially in extreme noisy conditions such as cocktail party. A limited number of research developments in the field of speech enhancement have been implemented into commercially
available hearing-aids. For example, spectral subtraction can be very effective in stationary conditions, but the processed speech remains unintelligible. In case of multiple microphones availability, beamforming algorithms can possibly lead to improvements in speech intelligibility. However, such approaches are hard to employ in an unpredictable noisy situations. Consequently, existing audio-only hearing aid approaches achieve  benefit just by simply amplifying signal and offering very little advantage for speech in high levels of noise \cite{hussaintowards}. 

Human performance in noisy environment is known to be dependent upon both aural and visual cues, which are combined by sophisticated multi-level integration strategies to improve intelligibility. The multimodal nature of the speech is well established  in literature, and it is well understood how speech is produced by the vibration of vocal folds and configuration of the articulatory organs. The correlation between the visible properties of the articulatory organs (e.g., lips, teeth, tongue) and speech reception has been previously shown in numerous behavioural studies \cite{sumby1954visual}\cite{summerfield1979use}\cite{mcgurk1976hearing}\cite{patterson2003two}. Therefore, a clear visibility of some articulatory organs could be effectively utilized to extract a clean speech signal out of a noisy audio background. The biggest advantage of using visual cues to extract clean audio features is their inherent noise immunity \cite{almajai2011visually}.  

However, embracing the multimodal nature of speech presents both opportunities and challenges for hearing assitive technology. The real-time implementation of AV hearing-aid demands for high datarate, low latency, low computational complexity, and high security. To address these challenges, we propose a novel integration of 5G, IoT, and strong privacy algorithms. The proposed AV hearing-aid framework is envisioned to address challenges such as cybersecurity attacks (location privacy, eavesdropping), interference between medical IoT devices (that can cause hearing-aids to operate incorrectly with potentially life threatening consequences), low-cost wireless technology design, low power consumption, limited battery, and high datarate requirement. Inspired by our previous work \cite{adeel2016random}, the novel wireless hearing-aid framework offloads the computational complexity and real-time optimization issues by running deep learning and big data optimization processes in the background on the cloud. The hearing-aid transmits the encrypted compressed audio/visual information and receives encrypted enhanced reconstructed speech in real-time. The hearing-aid connects to an indoor 5G wireless access point and back/fronthaul core network that serves as the communication infrastructure of the system. 

The rest of the paper is organized as follows: Section 2 presents the envisioned 5G IoT enabled secure AV-hearing aid framework. Section 3 presents the proposed real-time lightweight chaotic encryption algorithms including applied transformations, image, and audio encryption flow diagrams. Section 4 explains the proposed speech enhancement framework including designed enhanced visually derived Wiener filter (EVWF) and long-short term memory (LSTM) based lip-reading regression model. In Section 5, the used AV datasets and feature extraction methodologies are presented. Section 6 presents the performance evaluation of the proposed AV encryption and speech enhancement algorithms. Finally, Section 6 concludes this work.

\section{5G IoT enabled Secure Audio-Visual Hearing-Aid Framework}
\label{sec:headings}

Modern digital hearing aids are marvels of sophisticated engineering. To hear modern audio, a low-latency and high datarate wireless solution is needed that would enable in-ear
hearing devices to connect seamlessly \cite{einhorn2017hearing}. In this article, a novel 5G IoT enabled secure audiovisual hearing-aid framework is proposed to acquire desired high quality processed speech in noisy environments. An example of state-of-the-art 5G IoT architecture is shown in Figure 1. The IoT enabled devices, supporting a wide variety of applications, to connect to the Internet, utilizing gateway for connectivity. For gateway design, different access technologies such as WiFi and 4G LTE could be used. However, they both are incapable of supporting thousands of connected IoT devices. WiFi suffers from packet collision and limited quality of service (QoS), whereas 4G LTE suffers from high delay and high packet loss for large number of users \cite{saxena2017efficient}. In addition, the wireless systems operating in unlicensed frequency bands require additional network equipment, resulting extra operation and capital expenditures. The unlicensed solutions are also prone to congestion with exponential increase in IoT deployment. In contrast, the next generation 5G wireless networks \cite{agiwal2016next}\cite{andrews2014will} are capable of providing higher datarates, enhanced mobile coverage, improved user experience at relatively lower cost, and dense connectivity \cite{bhushan2014network}. Furthermore, to address aggravating detrimental greenhouse (CO2) gas emissions due to ultra-dense 5G wireless networks  and increased network’s energy consumption, 5G C-RAN is a widely accepted solution that enables improved environmental sustainability, OPEX, resource management, and energy efficiency \cite{saxena2017efficient}.
\begin{figure}[h]
	\centering
	\includegraphics[width=0.60\textwidth]{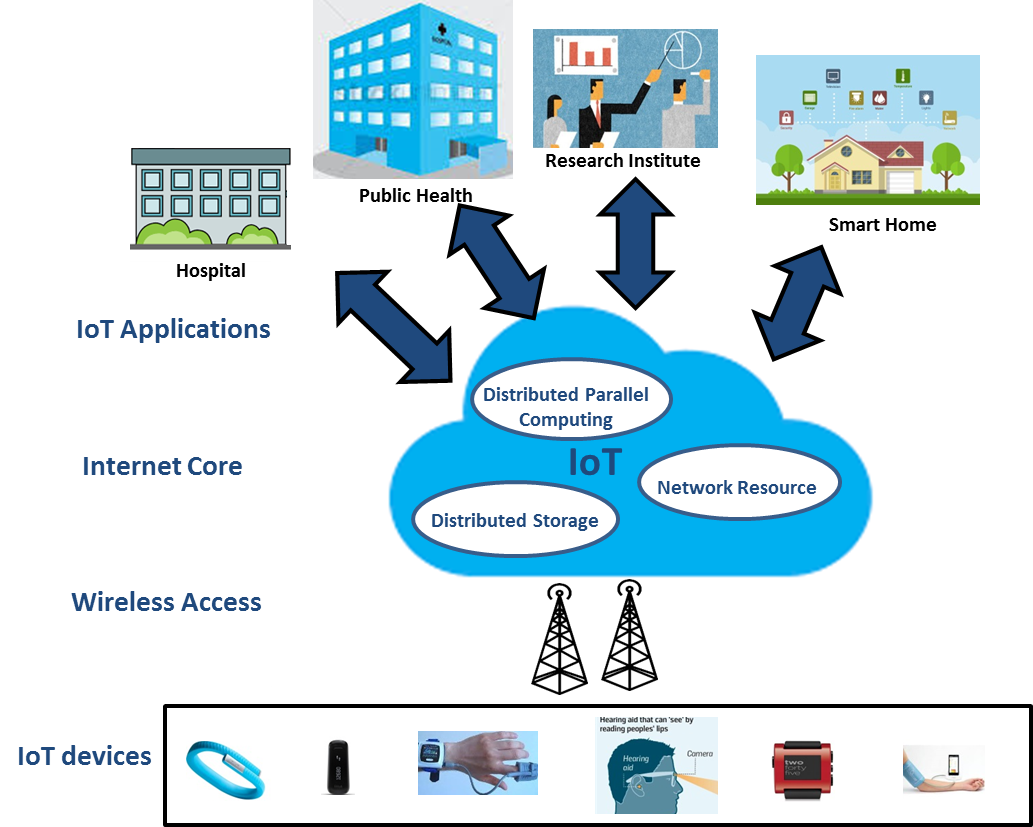} 
	\caption{5G IoT Architecture}
	\label{a}
\end{figure}

\begin{figure}[!htb]
	\centering
	\includegraphics[width=0.85\textwidth]{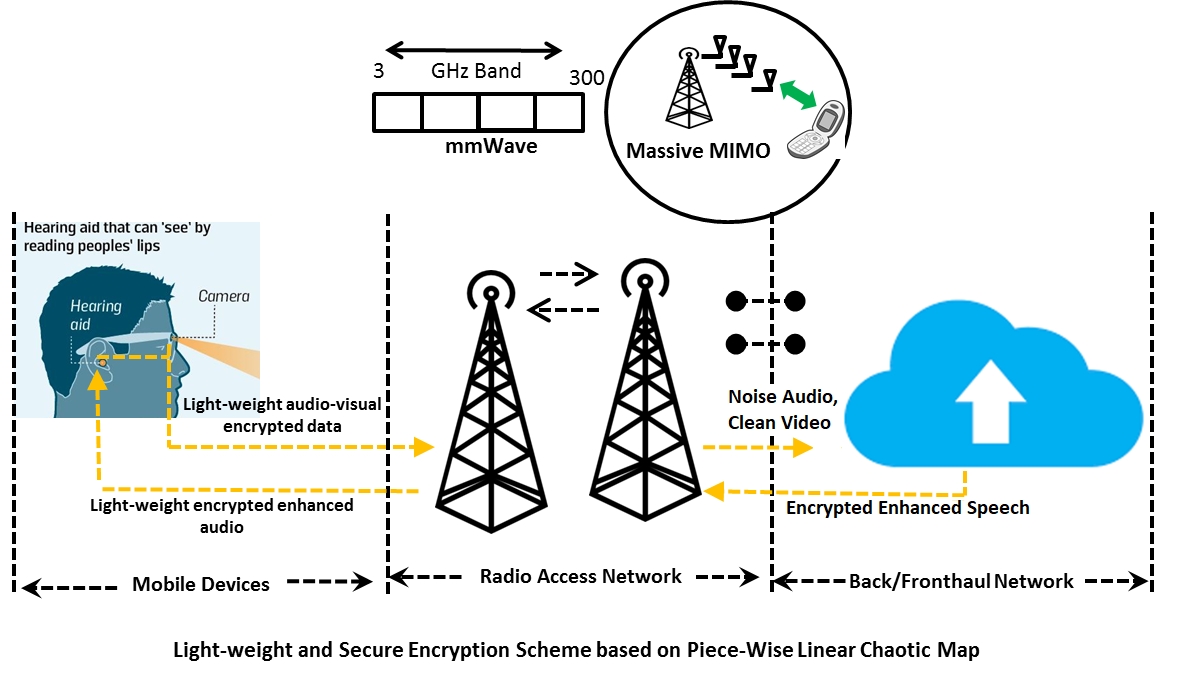} 
	\caption{Envisioned 5G IoT enabled Secure Audio-Visual Hearing-Aid}
	\label{a}
\end{figure}
 \begin{figure*}
	\centering
	\includegraphics[trim=0cm 0cm 0cm 0cm, clip=true, width=0.75\textwidth]{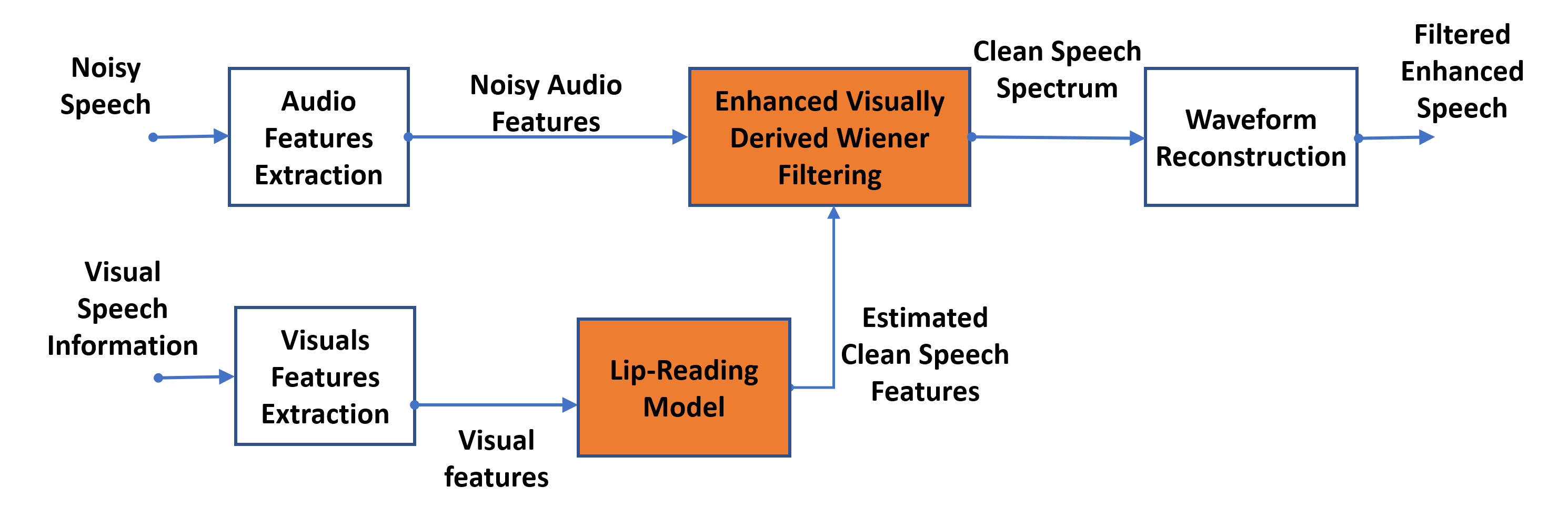}
	\caption{Proposed Lip-Reading Driven Deep Learning Approach for Speech Enhancement}
	\label{fig:Picture2}
\end{figure*}


The envisioned 5G IoT enabled secure audiovisual hearing-aid is depicted in Figure 2. It is to be noted that the computational complexity and real-time processing issues due to deep learning and big data optimization processes are addressed by running them in the background on the cloud. The mobile hearing-aid only transmits the encrypted compressed AV information and receives encrypted enhanced reconstructed speech in real-time. For end-to-end communication, an indoor 5G wireless small sized cell and back/fronthaul core network are proposed as the communication infrastructure of the system \cite{chen20175g}. For IoT gateway, we propose the use of an efficient IoT gateway over 5G wireless system (i.e. an integration of Massive MIMO, mmWave, 5G-Cloud-RAN (C-RAN), and indoor small sized cells for line of sight (LOS) signalling), developed and tested in \cite{saxena2017efficient}. The developed IoT gateway in \cite{saxena2017efficient} promises the uplink latency of 10ms and 5ms with and without compression respectively. In addition, it ensures minimum interference between medical IoT devices with low power consumption. The novelty of these gateways lie in efficient uplink IoT traffic classification and optimal uplink data (traffic) compression strategies. This helps in relaxation of uplink traffic burden and results in efficient utilization of uplink wireless resources. More details on 5G-CRAN and front/backhaul connectivity are comprehensively presented in \cite{saxena2017efficient}. 

For real-time  lightweight  audio-visual encryption, PWLCM,  Chebyshev  map,  secure  hash  algorithm and  a  novel  S-Box algorithms  are utilized. In the literature, conventional encryption approaches such as advanced encryption standard (AES) and Rivest–Shamir–Adleman (RSA)/Elliptic Curve (signing) are suitable for high processing power systems but incompatible with embedded low power sensor networks. Therefore, lightweight cryptography can potentially address real-time encryption challenges \cite{buchanan2017lightweight}. In our proposed scheme, the encrypted audio and video signals are exploited in the cloud by the designed novel lip-reading driven speech enhancement system, depicted in Figure 3. The proposed speech enhancement approach leverages the complementary strengths of both deep learning and analytical acoustic modelling (filtering based approach) that operates at two levels. In the first level, a novel deep learning based lip-reading regression model is employed. In the second level, lip-reading approximated clean-audio features are exploited, using an EVWF, for estimating the clean audio power spectrum. Finally, the Wiener filter is applied to the magnitude spectrum of the noisy input audio signal, followed by the inverse fast Fourier transform (IFFT), overlap, and combining processes to produce enhanced magnitude spectrum. More details are presented in Section 4. The proposed AV speech enhancement framework finally transmits the enhanced encrypted speech to the mobile hearing-aid. 

\section{Proposed Real-Time Lightweight Chaotic Encryption} \label{Prel}
In the proposed scheme, PWLCM, Chebyshev map, SHA and a novel S-Box algorithms are effectively used for real-time light-weight encryption. The applied transformations are briefly explained in the subsequent sections.

\subsection{Applied Transformations}
\subsubsection{PWLCM}
As outlined in shannon novel paper \cite{shannon1949communication}, a good encryption scheme is composed of two stages: (i) Confusion and (ii) Diffusion. In confusion stage, a correlation between key and ciphertext is made complex. Diffusion means that a minor change in plaintext should change the corresponding ciphertext significantly. The proposed algorithm uses PWLCM in confusion process. The PWLCM can be written as: 

\begin{equation} \label{tent}
y_{n+1} = f(y_{n},\lambda)=
\begin{cases}
\frac{y_{n}}{\lambda}, & $if $  y_{n} \in [0,\lambda] \\
\frac{1-y_{n}}{1-\lambda}, & $if $ y_{n} \in (\lambda,0.5] \\
F(1-y_{n}), & $if $ y_{n} \in (0.5,1], \\
\end{cases}
\end{equation}

where, $y_{n}$ are pseudo-random chaotic values,  $y_{n} \in (0,1)$, and $\lambda$ is the control parameter. Both $\lambda$ and $y_{0}$ serve as an initial condition and called as key for chaotic pseudo-random number generation. 

\subsubsection{Chebyshev Map}
In \cite{huang2012image}, Huang et al., proposed a novel key generator method using Chebyshev map. The Chebyshev map can be defined mathematically as \cite{huang2012image,wang2014cryptanalysis}: 
\begin{equation}
T_k(x) =cos(k \times arc \mbox{ }cos(x)), 
\end{equation}
where \textit{k = 0, 1, 2, ...,N} and $x \in [-1,1]$. Huang suggested $k = 4$ for less computation and better use of Chebyshev. In the proposed scheme, k = 4 is used which defines the Chebyshev function as: 
\begin{equation}
f(x_i) =8 x^{4}_{i-1} - 8 x^{2}_{i-1} + 1, i = 1,2, ....N 
\end{equation}
\subsubsection{Logistic-Sine Map}
To overcome the drawbacks of one-dimensional (1D) Logistic map, Zhou et al. in \cite{zhou2014new} proposed a novel method of chaotic maps combination. For a larger chaotic map, the authors combined two exiting 1D Logistic and Sine maps. Logistic-sine map has many advantages over traditional maps and is mathematically defined as \cite{zhou2014new}:
\begin{equation}
z_{n+1} = (rz_n(1-z_n) + (4-r) \frac{sin(\pi z_n)}{4}) mod(1), 
\end{equation}

\subsubsection{Secure Hash Algorithm (SHA)}
SHA generates a fix length value known as hash code by applying some function to the plaintext message. In the literature, SHA has different variants depending on the size of the output e.g, SHA-1, SHA-256 and SHA-512 for 128, 256, and 512 bits outputs, respectively. In the proposed scheme, we used SHA-512 such that $H(m)$ = $h$(512 bits). Secret Key in the proposed scheme is dependent on SHA-512. A minor change in the plaintext generates a completely different hash and different initial key parameters. 
\subsubsection{Affine Transformation}
Affine transformation is a one to one mapping that transforms a unique plaintext into a unique symbol. The following affine transformation is used in the proposed scheme:

\begin{equation} 
AT(w)= 
\begin{bmatrix} 
1  & 1 & 1  & 1 & 1  & 0 & 0  & 0 \\
0  & 1 & 1 & 1 & 1 & 1 &0  & 0 \\
0  & 0 & 1 & 1 & 1 & 1 & 1  & 0 \\
0 & 0 & 0  & 1 & 1  & 1 & 1 & 1 \\
1  & 0 & 0  & 0 & 1  & 1 & 1  & 1 \\
1  & 1 & 0  & 0 & 0  & 1 & 1  & 1 \\
1  & 1 & 1  & 0 & 0  & 0 & 1  & 1 \\
1  & 1 & 1  & 1 & 0  & 0 & 0  & 1 \\
\end{bmatrix} \times 
\begin{bmatrix}
w_{7}\\
w_{6}\\
w_{5}\\
w_{4}\\
w_{3}\\
w_{2}\\
w_{1}\\
w_{0}\\
\end{bmatrix} \oplus
\begin{bmatrix} 
0\\
1 \\
1\\
0 \\
0\\
0 \\
1\\
1 \\
\end{bmatrix}
\end{equation}
where $w_i$ are coefficients of $w$ i.e multiplicative inverse modulo ($w^8 +w^4+ w^2 + w^1 +1)$.  
\subsection{Image Encryption Scheme}
The flow chart for the grayscale images using PWLCM, Chebyshev, SHA-512 and affine transformation is shown in Figure 4. The detailed steps of our proposed cryptosystem are as follows: 
\begin{figure*}
	\center
	\includegraphics[width=0.5\textwidth]{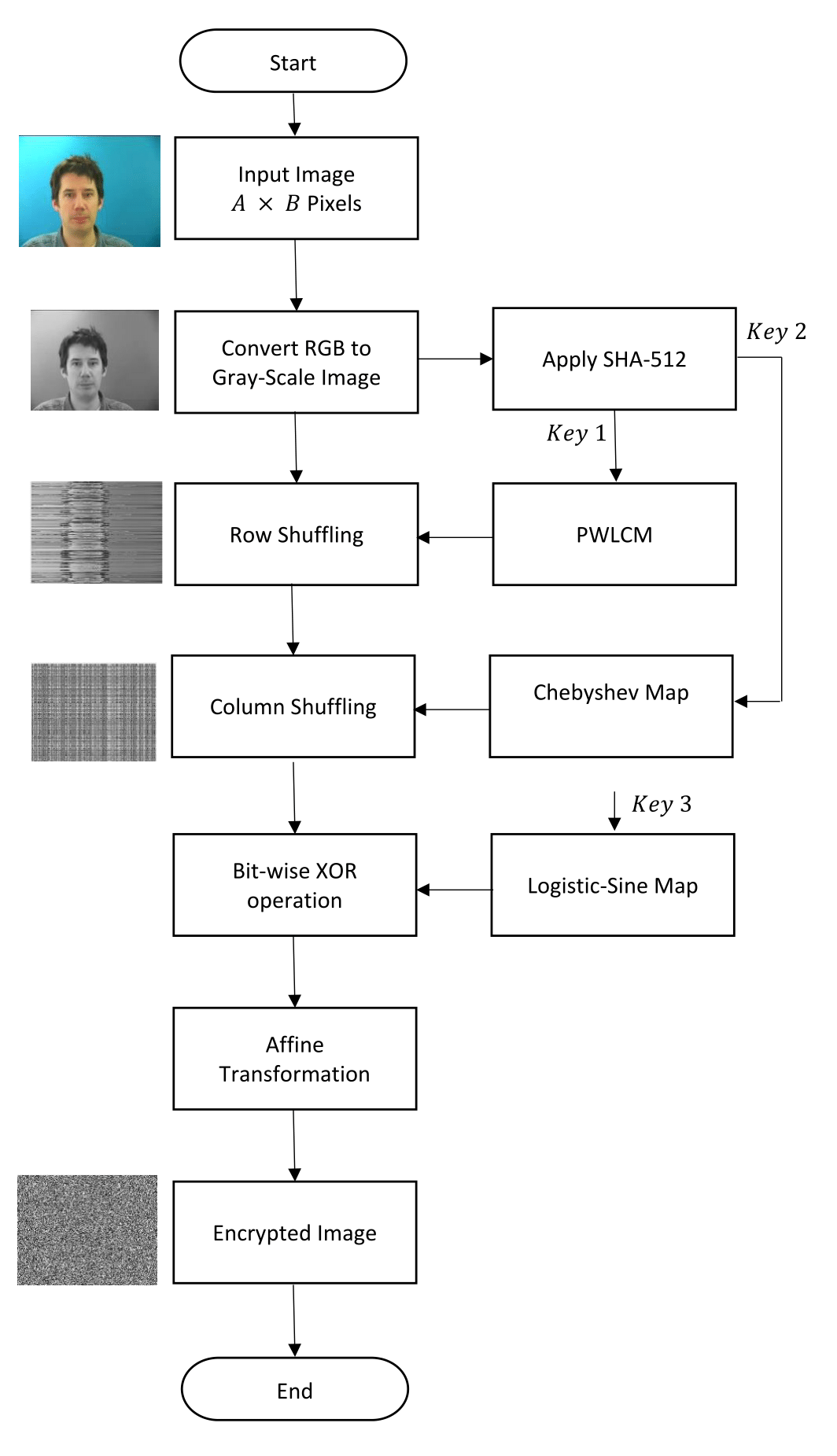}
	\caption{Proposed image encryption algorithm.}
	\label{flow}       
\end{figure*}

\begin{itemize}

	\item \textbf{Step 1}: Convert colour image $I_c$ of size $A \times B$ to gray-scale image $I_g$ and save result in $\psi$. 
	\item \textbf{Step 2}: Apply SHA-512 on gray-scale plaintext image $\psi$ and save hexadecimal hash value in variable $\theta$. 
	\item \textbf{Step 3}: Select first and last 12 hash values and save in $\kappa_1$ and  $\kappa_2$.
	\item \textbf{Step 4}: Convert hexadecimal values saved in $\kappa_1$ and $\alpha_1$ to decimal values and store result in $\kappa_1$, and $\beta_2$, respectively. 
	\item \textbf{Step 5}: Generate SHA-based initial conditions for PWLCM and Chebyshev using following equations:
	\begin{equation}
	y_0 = \dfrac{\kappa_1}{2^{48}}
	\end{equation}
	\begin{equation}
	x_0 = \dfrac{\kappa_2}{2^{48}}
	\end{equation}
	
	\item \textbf{Step 6}: Iterate PWLCM $A$ times and store chaotic values in $\alpha$. Randomly permute rows of gray-scale image $I_g$ using the sequence $\alpha$ and save values in $I_{rp}$. 
	
	\item \textbf{Step 7}: Iterate Chebyshev map $B$ times and store chaotic values in $\beta$. Randomly permute columns of $I_{rp}$ using the sequence $\alpha$ and save values in $I_{permuted}$
	
	\item \textbf{Step 8}: Iterate Logistic-sine map $A \times B$ times and store random values in $\gamma$. 
	\item \textbf{Step 9}: Apply following operations on $\gamma$:
	\begin{equation}
	R_1 = \mbox{Mod}(\gamma \times 10^{14}, 256),
	\end{equation}
	\begin{equation}
	R_2 = floor(R_1).
	\end{equation}
	
	\item \textbf{Step 10}: Rearrange row-vector $R_2$ in matrix form $R$ and Bit-wise XOR random matrix $R$ with $I_{permuted}$ to get $\phi$. 
	
	\item \textbf{Step 11}: Apply affine transformation on $\phi$ and store values as a ciphertext image $C$. 
\end{itemize}

For decryption, encryption steps are followed in the reverse order. 

\subsection{Audio Encryption Scheme}
The flow chart for the audio signal using PWLCM, Chebyshev, SHA-512 and affine transformation is shown in Figure \ref{flow}. The detailed steps of our proposed cryptosystem are as follows: 
\begin{figure*}
	\center
	\includegraphics[width=0.5\textwidth]{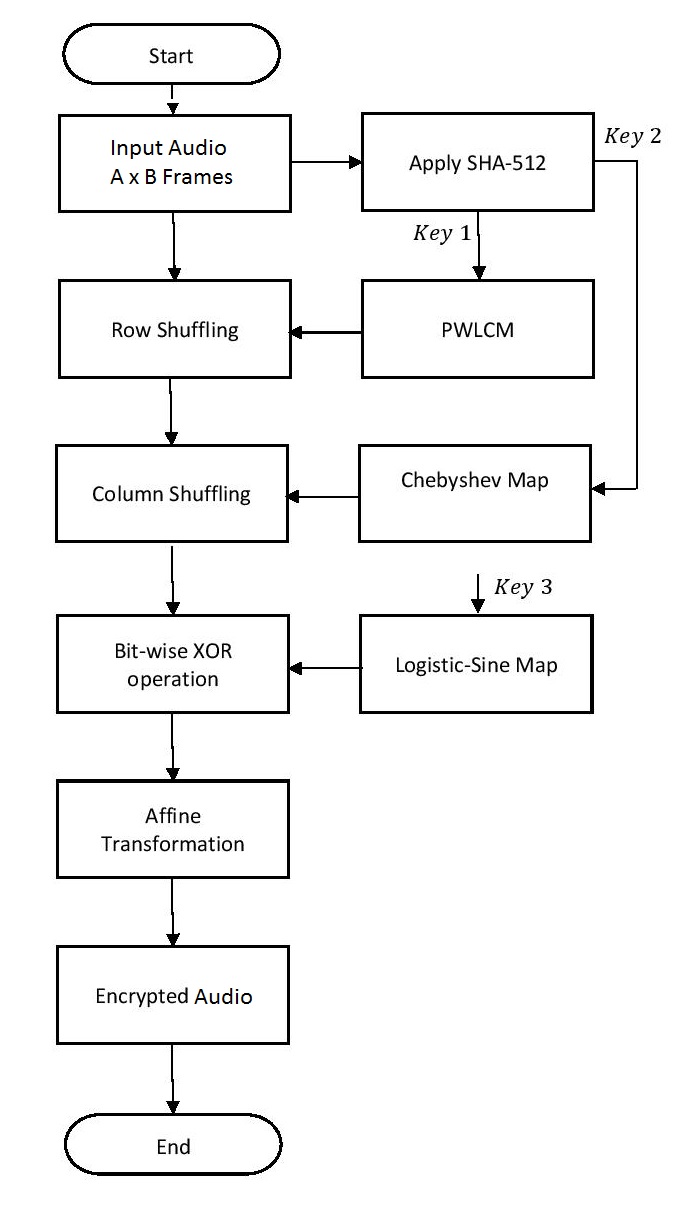}
	\caption{Proposed audio encryption algorithm.}
	\label{flow}       
\end{figure*}

\begin{itemize}
	
	\item \textbf{Step 1}: Convert 1D audio signal into 2D of size $A \times B$ and save result in $\psi$. 
	\item \textbf{Step 2}: Apply SHA-512 on $\psi$ and save hexadecimal hash value in variable $\theta$. 
	\item \textbf{Step 3}: Select first and last 12 hash values and save in $\kappa_1$ and  $\kappa_2$.
	\item \textbf{Step 4}: Convert hexadecimal values saved in $\kappa_1$ and $\alpha_1$ to decimal values and store results in $\kappa_1$, and $\beta_2$, respectively. 
	\item \textbf{Step 5}: Generate SHA-based initial conditions for PWLCM and Chebyshev using equations (6-7)
	\item \textbf{Step 6}: Iterate PWLCM $A$ times and store chaotic values in $\alpha$. Randomly permute rows of audio vector $A_g$ using the sequence $\alpha$ and save values in $I_{rp}$. 
	\item \textbf{Step 7}: Iterate Chebyshev map $B$ times and store chaotic values in $\beta$. Randomly permute columns of $A_{rp}$ using the sequence $\alpha$ and save values in $A_{permuted}$
	
	\item \textbf{Step 8}: Iterate Logistic-sine map $A \times B$ times and store random values in $\gamma$. 
	\item \textbf{Step 9}: Apply operations given in equations (8-9) to $\gamma$.
	\item \textbf{Step 10}: Rearrange row-vector $R_2$ in matrix form $R$ and Bit-wise XOR random matrix $R$ with $A_{permuted}$ to get $\phi$. 
	\item \textbf{Step 11}: Apply affine transformation on $\phi$ and store values as a ciphertext audio $C$. 
\end{itemize}

For decryption, encryption steps are followed in the reverse order.

\section{Speech Enhancement Framework} 
The state-of-the-art VWF and designed EVWF are depicted in Figure 6 (a) and (b) respectively. The authors in \cite{almajai2011visually} presented a hidden Markov model-Gaussian mixture model (HMM/GMM) based two-level state-of-the-art VWF for speech enhancement. However, the use of HMM/GMM models for the estimation of clean audio features from visual features and cubic spline interpolation for the approximation of high dimensional clean audio power spectrum from the estimated low dimensional audio features are not optimal choices. The HMM/GMM model suffers from poor generalization and cubic spline interpolation method fails to estimate the missing power spectral values that leads to a poor audio power spectrum estimation. In contrast, the designed EVWF addressed the limitations of state-of-the-art VWF \cite{almajai2011visually} by employing an inverse filter-bank transformation (i.e. a pseudoinverse of the approximated audio features) for audio power spectrum estimation as compared to the cubic spline interpolation method. In addition, the use of LSTM addressed the generalization and accurate clean speech coefficient estimation issues. The designed EVWF also eliminates the need for voice activity detection (VAD) and noise estimation. The designed LSTM based lip-reading model consists of input layer, two LSTM layers, and output dense layer. In the designed LSTM model, prior visual features were feeded into the stacked LSTM layers to exploit the existing temporal correlation. The lower LSTM layer used 250 cells for encoding the input visual information and passed its hidden state to the second LSTM layer, which has 300 cells. The output of the second LSTM layer was then feeded into the fully connected (dense) layer which has total 23 neurons with linear activation function. The designed LSTM model was trained with the objective to minimise the mean squared error (MSE) between the predicted and the actual audio features using stochastic gradient decent algorithm and RMSProp optimiser. More details on EVWF mathematical formulation and LSTM model are comprehensively presented in our previous work \cite{adeel2018lip}. 
\begin{figure}
	\centering 
	\subfloat[State-of-the-art visually-derived Wiener filtering \cite{almajai2011visually}. ]{\includegraphics[width=0.5\textwidth]{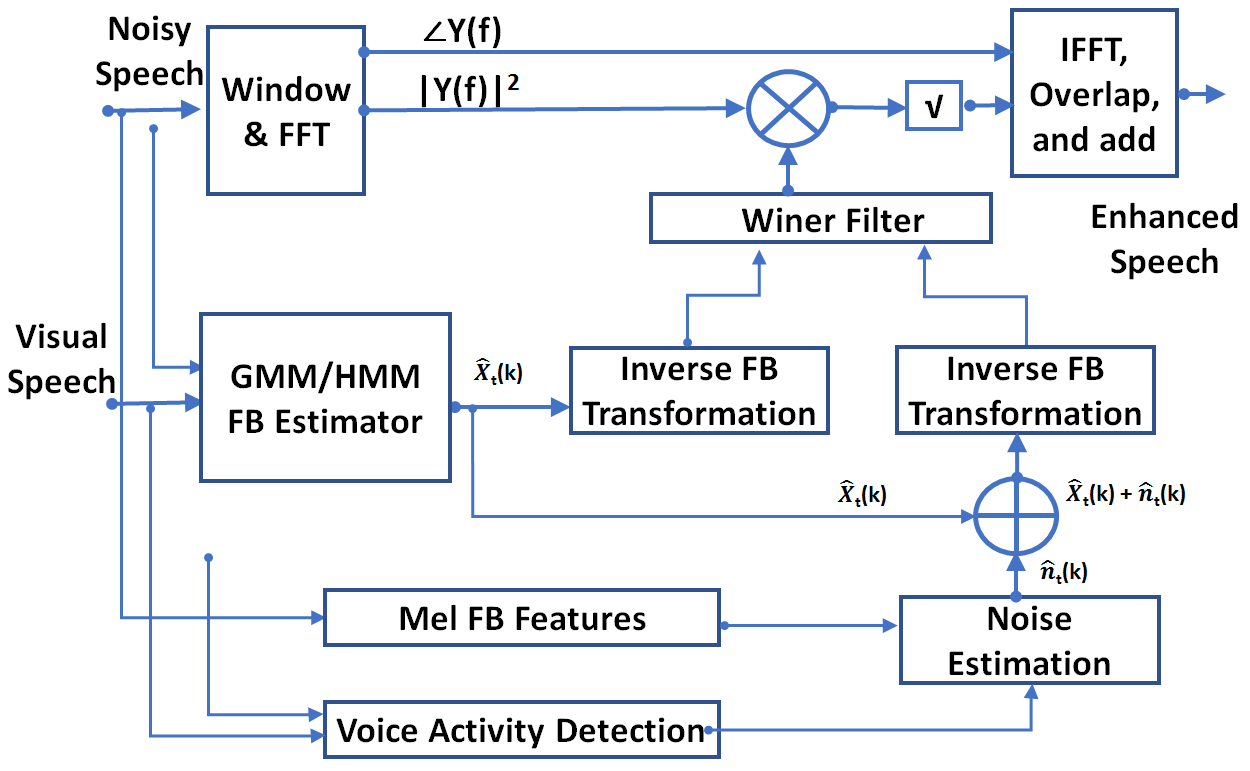}} 
	\subfloat[Proposed enhanced visually-derived Wiener filtering]{\includegraphics[width=0.5\textwidth]{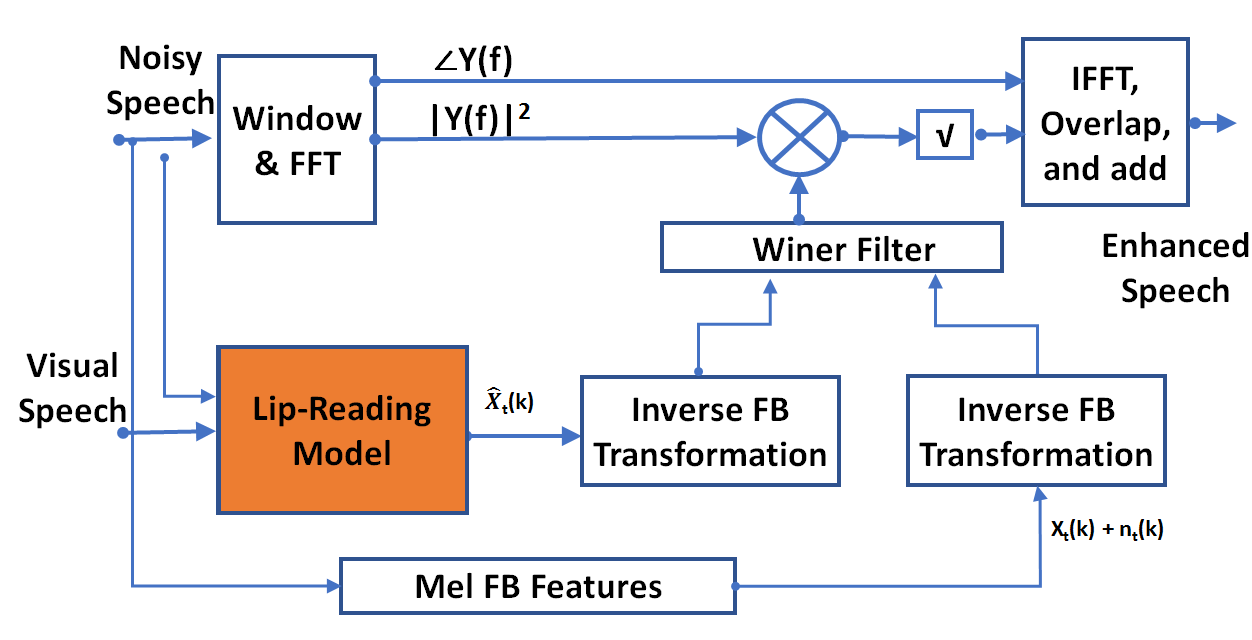}}\\
	%
	
	\caption{State-of-the-art visually-derived Wiener filtering and proposed enhanced visually-derived Wiener filtering} 
	\label{fig:inputport} 
\end{figure}

\section{Dataset and Audio-Visual Feature Extraction}
For AV encryption and speech enhancement, Grid \cite{cooke2006audio} and ChiME3 \cite{barker2015third} corpora are used. For noisy utterances generation, the clean videos from Grid corpus were mixed with ChiME3 noises for different SNR levels ranging from -12 to 12dB. For preprocessing, sentence alignment and prior visual frames were used to stop the model from learning redundant information and improve mapping between visual and audio features (exploiting their temporal information).  
\subsection{Audio-Visual feature extraction}
For audio feature extraction, the input audio signal was sampled at 50kHz and segmented into \textit{N} 16ms frames with 800 samples per frame and 62.5\% increment rate. To produce 2048-bin power spectrum, a hamming window and Fourier transformation was applied, followed by the logarithmic compression to produce 23-D log-FB signal. The visual features were extracted from the Grid Corpus videos recorded at 25 fps. The video files were processed by extracting a sequence of individual frames and applying a Viola-Jones lip detector \cite{viola2001rapid} and object tracker \cite{ross2008incremental}. Furthermore, to ensure appropriate lip tracking, processed utterances were manually validated. Finally, the 2D-Discrete Cosine Transformation (2D-DCT) was applied to produce vectors of pixel intensities, followed by interpolation. More details are presented in our previous work \cite{adeel2018lip}. 
\section{Performance Evaluation}
\begin{figure*}
	\centering     
	\subfloat[Original Image]{\includegraphics[width=0.4\textwidth]{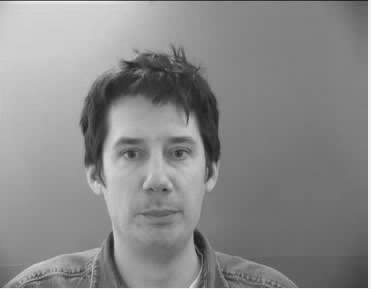} }\hspace*{0.5cm}                
	\subfloat[Encrypted Image]{\includegraphics[width=0.4\textwidth]{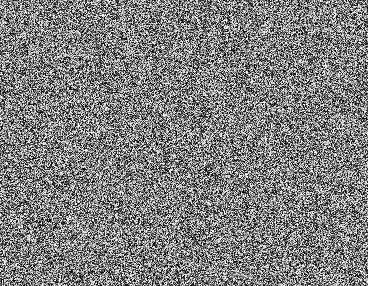}} \hspace*{0.5cm} \\
	\caption{Encryption Results. It is to be noted that the proposed encryption scheme completely concealed the plaintext information.}
	\label{plaintext}
\end{figure*} 
\begin{figure*}
	\centering     
	\subfloat[Original Image Histogram.  ]{\includegraphics[width=0.4\textwidth]{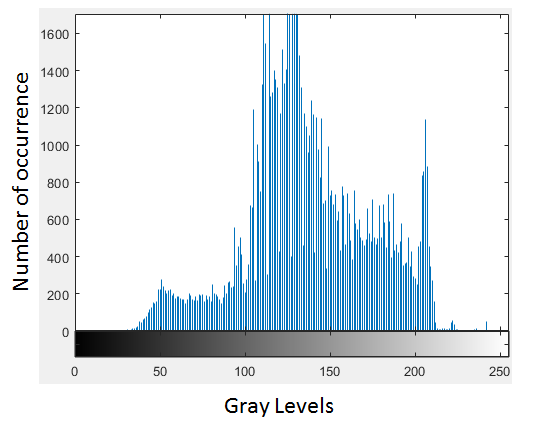}}\hspace*{0.75cm}                
	\subfloat[Encrypted Image Histogram. ]{\includegraphics[width=0.4\textwidth]{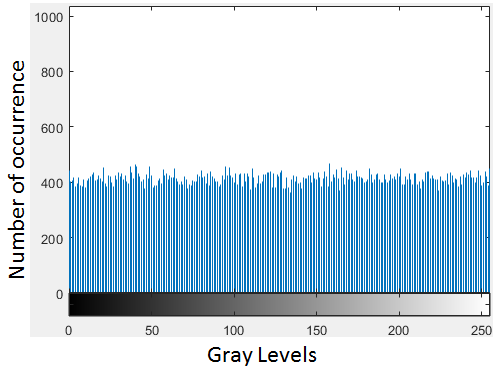}} \hspace*{0.75cm} \\
	\caption{Histogram Results. It is to be noted that the obtained histogram of the encrypted image is flat which is ideally required.}
	\label{hist}
\end{figure*} 
\subsection{Lip-Images Encryption and Security Analyses}
\begin{figure*}
	\centering     
	\subfloat[Original Image correlation coefficient.  ]{\includegraphics[width=0.4\textwidth]{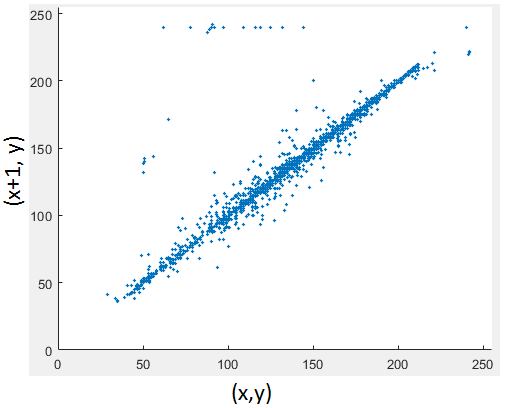} }\hspace*{0.5cm}                
	\subfloat[Encrypted Image correlation coefficient. ]{\includegraphics[width=0.4\textwidth]{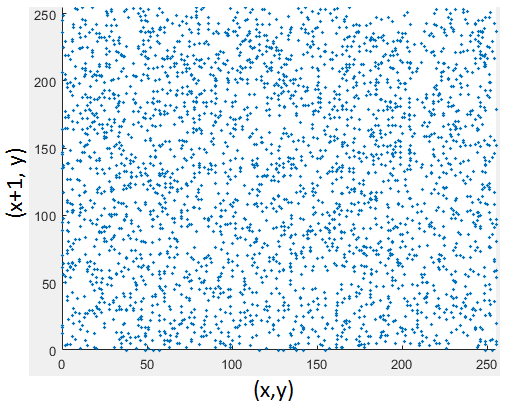}  } \hspace*{0.5cm} \\
	
	\caption{Correlation coefficients in vertical direction.}
	\label{corr}
\end{figure*}

In order to show the effectiveness of the encryption scheme, one test image is selected with a specific lips position. The proposed lightweight encryption scheme is applied to the plaintext image and the obtained results are shown in Figure 7. It can be seen that the proposed scheme completely concealed the plaintext information. Moreover, the histogram results shown in Figure \ref{hist} acquired the flat histogram which is required ideally. However, the histogram results are not sufficient to prove a complete security of the cryptosystem. Therefore, a large number of security metrics such as correlation coefficient, entropy, contrast, energy,  NPCR, and UACI defined in our previous work \cite{ahmad2017compression,khan2017novel,khan2017td} are used.  The degree of similarity between adjacent pixels are generally analysed via correlation coefficient metric. Ideally, correlation in all directions (horizontal $(H_{CC})$, vertical$ (V_{CC})$ and diagonal $(D_{CC})$) should be close to zero. Entropy is an other important metric which can evaluate the resistance capability against statistical attacks. For a good cryptosystem, entropy of a gray-scale encrypted image should be 8 bits ideally. A contrast of an encrypted image is defined as intensity between a pixel and its neighbour pixels. In image encryption, higher values of contrast indicates higher quality of encrypted image. A sum of squared elements (SSE) in gray level co-occurrence matrix returns energy of an image. The energy value of a secure encrypted image is desired to be low. In case of complete constant pixels, energy value is 1. NPCR and UACI show resistant against differential attacks. Higher values of NPCR and UACI reflects higher encryption quality. More detail on how these parameters prove the security of our encryption scheme is defined in detail in our previous work \cite{ahmad2017compression,khan2017novel,khan2017td}. \\
The correlation plots in vertical directions are shown in Figure \ref{corr}. From correlation plots, it is evident that the distribution of adjacent pixels in vertical direction is uncorrelated as compared to plaintext correlation. Mathematical values of correlations in vertical, horizontal and diagonal directions are outlined in Table 1. In the table, lower correlation values prove the robustness of the proposed scheme. In addition, it can be seen that the histogram plot for the proposed scheme is uniformly distributed; hence, assures resistance to statistical attacks as compared to existing algorithms \cite{ahmad2015chaos}\cite{anees2014chaotic}. NPCR  greater than 99\% also reveals higher security. Consequently, the security  metrics demonstrate the effectiveness and higher security of the proposed scheme. Lastly, the required encryption/decryption time is less than 25 msec on a 60 GB RAM using MATLAB software. Such a low processing time confirms that the proposed scheme is lightweight and could be an effective solution for practical real-time applications. However, further real-time optimization is an going work.

\begin{table}
	\center
	\setlength{\tabcolsep}{0.4em}
	\renewcommand{\arraystretch}{1}
	\caption{Image security assessment. It is to be noted that the correlation values in all directions (horizontal $(H_{CC})$, vertical$
		 (V_{CC})$ and diagonal $(D_{CC})$) are low which proves the robustness of the proposed scheme. In addition, note the resistance capability against statistical attacks and high security, evaluated using Contrast, Energy, NPCR and UACI tests.}
	\label{tab:image1}       
	\begin{tabular}{lll}
		\hline
		Security Parameter & Original Frame & Encrypted Frame \\
		\hline
		$V_{CC}$ & 0.9523 & 0.0044 \\
		$H_{CC}$ & 0.9711 & -0.0056 \\
		$D_{CC}$ & 0.9677 & 0.0089  \\
		$Entropy$  & 7.0025 & 7.9983 \\
		$Contrast$ & 0.1049  & 10.4608  \\
		$Energy$ & 0.2161 & 0.0156  \\
		$NPCR$  & NA & 99.4566 \\
		$UACI$  & NA & 33.1561 \\
		\hline
	\end{tabular}
\end{table}

\begin{table}
	\centering 
	\caption{LSTM Training and Testing Accuracy - Comparison For Different Visual Frames. The table presents an overall behaviour of the LSTM model when contextual information (i.e. previous frames) is added. It is to be noted that the LSTM model exploited the temporal correlation effectively but saturated at 18 prior visual frames.}  
	\begin{tabular}{|c|c|c|}    
		\cline{2-3}                      
		\multicolumn{1}{c}{} &  \multicolumn{2}{|c|}{LSTM} \\
		\hline
		Visual Frames &  $MSE_{train}$ & $MSE_{test}$ \\
		\hline
		\hline	
		1  & 0.092 & 0.104 \\
		2   & 0.087 & 0.097 \\
		4   & 0.073 & 0.085 \\
		8  & 0.066 & 0.082 \\
		14  & 0.061 & 0.080 \\		
		18  & 0.058 & 0.078 \\		
		\hline	 		
	\end{tabular}
	\label{table:featextract-sentnum}    
\end{table}
\subsection{Speech Enhancement Results}
\subsection{Lip-Reading Results}
For lip-reading, multiple prior visual frames (lip images) are used (ranging from 1 visual frame to 27 prior visual frames). The simulation results are shown in Table 2. It can be seen that by moving from 1 visual frame to 18 visual frames, a significant performance improvement could be achieved. The LSTM model with 1 visual frame achieved the MSE of 0.092, whereas with 18 visual frames the model achieved the least MSE of 0.058. LSTM based learning model exploited the temporal information (i.e. prior visual frames) efficiently and showed consistent reduction in MSE while going from 1 to 18 visual frames. This is mainly because of its inherent recurrent architectural property and the ability of retaining state over long time spans by using gates. More details, critical analysis, and comparisons are comprehensively presented in our previous work \cite{adeel2018lip}. 
 \begin{table*}[t]
 	\caption{Speech Enhancement Results. It can be seen that at low SNR levels, EVWF significantly outperformed benchmark SS and
 		LMMSE based speech enhancement methods.}
 	\label{tab:tfaccuracy}
 	\centering
 	\resizebox{0.6\textwidth}{!}{%
 		\begin{tabular}{c|ccc|ccc}
 			\toprule
 			\multicolumn{1}{c}{\textbf{SNR}} &
 			\multicolumn{3}{c|}{\textbf{PESQ}} &
 			\multicolumn{3}{c}{\textbf{MOS}}\\
 			\midrule
 			&
 			\multicolumn{1}{c}{\textbf{SS}} &
 			\multicolumn{1}{c}{\textbf{LMMSE}} &
 			\multicolumn{1}{c|}{\textbf{EVWF}} &
 			\multicolumn{1}{c}{\textbf{SS}} &
 			\multicolumn{1}{c}{\textbf{LMMSE}} &
 			\multicolumn{1}{c}{\textbf{EVWF}}
 			
 			\\
 			\midrule
 			-12dB & 0.9 &	0.95	& 1.52 &	 0.31 &	0.315 &	1.68 \\  
 			-6dB & 1.01&	1.03	&1.58& 0.51& 	0.48& 	1.965 \\
 			-3dB & 1.17&	1.18&	1.60	& 	1.17& 	1.165& 	2 \\
 			0dB & 		1.21&	1.20&	1.63 &		1.925 & 	1.97& 	2.11 \\
 			3dB &	1.25&	1.34&	1.72	& 2.025 &	2.085& 	2.22  \\
 			6dB &1.26&	1.39&	1.69	& 	2.295& 	2.345& 	2.33 	\\
 			12dB& 1.54&	1.60&	1.74&		2.58& 	2.61& 	2.54 	\\
 			
 			\bottomrule
 		\end{tabular}
 	}
 \end{table*}

\subsubsection{Objective Test}
For objective testing and comparison with the state-of-the-art audio only speech enhancement methods (spectral subtraction (SS) and Log-Minimum Mean Square Error (LMMSE)), perceptual evaluation of speech quality (PESQ) is used to evaluate the quality of restored speech. PESQ is one of the reliable methods to evaluate speech quality. The PESQ score is computed as a linear combination of the average disturbance value and the average asymmetrical disturbance values. The PESQ score ranges from -0.5 to 4.5 corresponding to low to high speech quality. The PESQ scores for proposed EVWF and state-of-the-art benchmark audio only speech enchantment approaches are depicted in Table 3. It can be seen that at low SNR levels, EVWF significantly outperformed both SS and LMMSE based speech enhancement methods.
\subsubsection{Subjective Listening Tests} 
The subjective listening test was conducted in terms of MOS with self-reported normal-hearing listeners. The listeners were presented with both clean (target) and enhanced speech, and were asked to rate the re-constructed speech on a scale of 1 to 5. The five rating choices were: (5) Excellent (when the listener feels unnoticeable difference compared to the target clean speech) (4) Good (perceptible but not annoying) (3) Fair (slightly annoying) (2) Poor (annoying), and (1) Bad (very annoying). The EVWF performance was compared with two state-of-the-art speech enhancement methods (SS and LMMSE). A total of 10 listeners took part in the evaluation session. In Table 3, it can be seen that at low SNRs (-12dB, -6dB, and -3dB), the proposed EVWF outperformed audio-only speech enhancement methods. On the other hand, for high SNRs, our AV approach performed comparably to the Audio only approach. 

 \begin{figure*}
 	\centering     
 	\subfloat[Original Audio]{\includegraphics[width=0.4\textwidth]{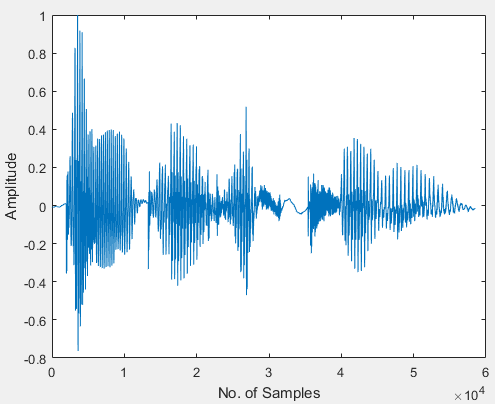} }\hspace*{0.5cm}                
 	\subfloat[Encrypted Audio]{\includegraphics[width=0.4\textwidth]{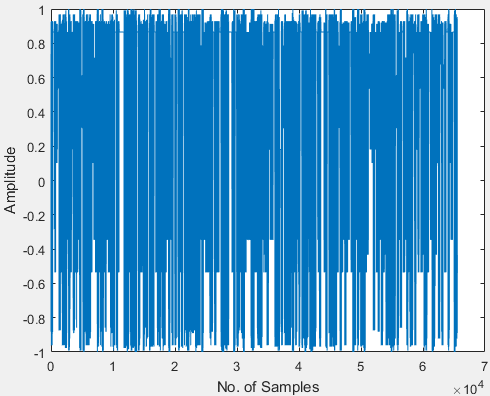}} \hspace*{0.5cm}\\
 	\subfloat[Decrypted Audio]{\includegraphics[width=0.4\textwidth]{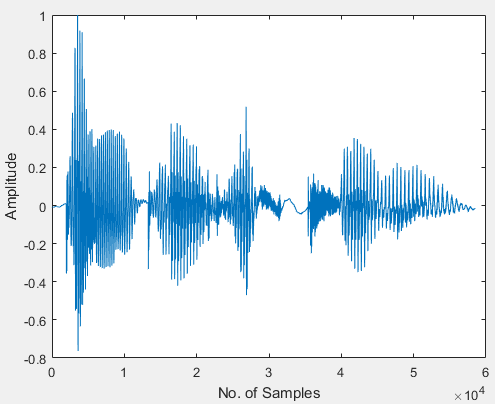}} \hspace*{0.5cm}
 	\caption{Audio encryption results.}
 	\label{plaintext}
 \end{figure*} 
 
 
%
\subsection{Audio Encryption and Security Analyses}
The proposed lightweight chaotic encryption was applied to the enhanced audio signal. A single bit in the original speech signal was modified and encrypted via same key \cite{sathiyamurthi2017speech}. The two ciphered speech signals, i.e., $S_c1$ and $S_c2$ were generated and the obtained results are shown in Figure 10. It can be seen that the proposed scheme completely concealed the audio information and decrypted speech signal accurately. Furthermore, to check the robustness and security strength of the proposed audio encryption scheme, security analysis tests such as number of sample change rate (NSCR), UACI, correlation coefficient, and key length are conducted, where low correlation value, 99.95\% NSCR, and 33.3\% UACI demonstrate the robustness of the proposed encryption scheme. In addition, the key length of the proposed approach is much greater than the minimum required length ($2^{100}$) that shows resistance against brute force attack. Table 4 presents the security analysis and effectiveness of the proposed scheme. 
%
%

\begin{table}
	\center
	\setlength{\tabcolsep}{0.4em}
	\renewcommand{\arraystretch}{1}
	\caption{Audio security assessment.}
	\label{tab:image1}       
	\begin{tabular}{ll}
		\hline
		Security Parameter & Encrypted Signal \\
		\hline
			NSCR  & 99.95\% \\
		UACI & 33.39\%  \\
		Correlation coefficient & 0.00022  \\
		Key length  & $10^{45}$ \\
		\hline
	\end{tabular}
\end{table}

\section{Conclusion}\label{conclusion}
The next-generation multimodal (lip-reading driven) hearing-aids stand as a major enabler for modern digital hearing aids, capable of restoring the intelligibility and reducing cognitive load in environments where overwhelming noise is present. However, the real-time implementation of such AV hearing-aids demand high data rate, low latency, low computational complexity, and high security.  
In this paper, we envisioned a 5G IoT enabled secure AV hearing-aid that leverages the complementary strengths of 5G and IoT technology to address the aforementioned challenges.  As part of the envisioned technology, two main contributions (AV speech enhancement in the cloud and lightweight AV encryption) are reported here. The critical analysis in terms of both speech enhancement and AV encryption demonstrate the potential of the envisioned technology in acquiring high quality speech reconstruction in extreme noisy situations and secure mobile AV hearing aid communication. Specifically, the comparative performance evaluation of the proposed speech enhancement method under real noisy environments revealed that the proposed approach significantly outperformed benchmark audio-only approaches at low SNR with comparable performance at high SNRs. The audio and video encryption results revealed the effectiveness of the proposed real-time lightweight encryption scheme in terms of both high security and processing time. The ongoing and future work includes the software integration of the proposed AV mobile hearing aid with 5G-CRAN and its hardware prototype implementation for real-time testing.

\section*{Acknowledgments}
This work was supported by the UK Engineering and Physical Sciences Research Council (EPSRC) Grant No. EP/M026981/1. The authors would like to gratefully acknowledge Mandar Gogate from the University of Stirling for his contribution in implementing LSTM driven AV mapping, which was published in our previous work and cited here for reference. 

\bibliographystyle{unsrt}
\bibliography{template.bib}

\begin{thebibliography}{10}

\bibitem{ruggles2012ignitability}
AJ~Ruggles and IW~Ekoto.
\newblock Ignitability and mixing of underexpanded hydrogen jets.
\newblock {\em international journal of hydrogen energy}, 37(22):17549--17560,
  2012.

\bibitem{kortlang2016combination}
S~Kortlang, S~Ewert, H~Meister, S~R{\"a}hlmann, J~Kie{\ss}ling, et~al.
\newblock Combination of controlled laboratory tests and structured field
  trials for a comprehensive evaluation of a model-based hearing aid.
\newblock {\em Int J Audiol}, 2016.

\bibitem{hussaintowards}
Amir Hussain, Jon Barker, Ricard Marxer, Ahsan Adeel, William Whitmer, Roger
  Watt, and Peter Derleth.
\newblock Towards multi-modal hearing aid design and evaluation in realistic
  audio-visual settings: Challenges and opportunities.

\bibitem{sumby1954visual}
William~H Sumby and Irwin Pollack.
\newblock Visual contribution to speech intelligibility in noise.
\newblock {\em The journal of the acoustical society of america},
  26(2):212--215, 1954.

\bibitem{summerfield1979use}
Quentin Summerfield.
\newblock Use of visual information for phonetic perception.
\newblock {\em Phonetica}, 36(4-5):314--331, 1979.

\bibitem{mcgurk1976hearing}
Harry McGurk and John MacDonald.
\newblock Hearing lips and seeing voices.
\newblock {\em Nature}, 264(5588):746, 1976.

\bibitem{patterson2003two}
Michelle~L Patterson and Janet~F Werker.
\newblock Two-month-old infants match phonetic information in lips and voice.
\newblock {\em Developmental Science}, 6(2):191--196, 2003.

\bibitem{almajai2011visually}
Ben Almajai, Milner.
\newblock Visually derived wiener filters for speech enhancement.
\newblock {\em IEEE Transactions on Audio, Speech, and Language Processing},
  19(6):1642--1651, 2011.

\bibitem{adeel2016random}
Ahsan Adeel, Hadi Larijani, and Ali Ahmadinia.
\newblock Random neural network based novel decision making framework for
  optimized and autonomous power control in lte uplink system.
\newblock {\em Physical Communication}, 19:106--117, 2016.

\bibitem{einhorn2017hearing}
Richard Einhorn.
\newblock Hearing aid technology for the 21st century: A proposal for universal
  wireless connectivity and improved sound quality.
\newblock {\em IEEE pulse}, 8(2):25--28, 2017.

\bibitem{saxena2017efficient}
Navrati Saxena, Abhishek Roy, Bharat~JR Sahu, and HanSeok Kim.
\newblock Efficient iot gateway over 5g wireless: A new design with prototype
  and implementation results.
\newblock {\em IEEE Communications Magazine}, 55(2):97--105, 2017.

\bibitem{agiwal2016next}
Mamta Agiwal, Abhishek Roy, and Navrati Saxena.
\newblock Next generation 5g wireless networks: A comprehensive survey.
\newblock {\em IEEE Communications Surveys \& Tutorials}, 18(3):1617--1655.

\bibitem{andrews2014will}
Jeffrey~G Andrews, Stefano Buzzi, Wan Choi, Stephen~V Hanly, Angel Lozano,
  Anthony~CK Soong, and Jianzhong~Charlie Zhang.
\newblock What will 5g be?
\newblock {\em IEEE Journal on selected areas in communications},
  32(6):1065--1082, 2014.

\bibitem{bhushan2014network}
Naga Bhushan, Junyi Li, Durga Malladi, Rob Gilmore, Dean Brenner, Aleksandar
  Damnjanovic, Ravi Sukhavasi, Chirag Patel, and Stefan Geirhofer.
\newblock Network densification: the dominant theme for wireless evolution into
  5g.
\newblock {\em IEEE Communications Magazine}, 52(2):82--89, 2014.

\bibitem{chen20175g}
Min Chen, Jun Yang, Yixue Hao, Shiwen Mao, and Kai Hwang.
\newblock A 5g cognitive system for healthcare.
\newblock {\em Big Data and Cognitive Computing}, 1(1):2, 2017.

\bibitem{buchanan2017lightweight}
William~J Buchanan, Shancang Li, and Rameez Asif.
\newblock Lightweight cryptography methods.
\newblock {\em Journal of Cyber Security Technology}, 1(3-4):187--201, 2017.

\bibitem{shannon1949communication}
Claude~E Shannon.
\newblock Communication theory of secrecy systems.
\newblock {\em Bell Labs Technical Journal}, 28(4):656--715, 1949.

\bibitem{huang2012image}
Xiaoling Huang.
\newblock Image encryption algorithm using chaotic chebyshev generator.
\newblock {\em Nonlinear Dynamics}, 67(4):2411--2417, 2012.

\bibitem{wang2014cryptanalysis}
Xingyuan Wang, Dapeng Luan, and Xuemei Bao.
\newblock Cryptanalysis of an image encryption algorithm using chebyshev
  generator.
\newblock {\em Digital Signal Processing}, 25:244--247, 2014.

\bibitem{zhou2014new}
Yicong Zhou, Long Bao, and CL~Philip Chen.
\newblock A new 1d chaotic system for image encryption.
\newblock {\em Signal processing}, 97:172--182, 2014.

\bibitem{adeel2018lip}
Hussain Amir Whitmer William~M Adeel~Ahsan, Gogate~Mandar.
\newblock Lip-reading driven deep learning approach for speech enhancement.
\newblock {\em arXiv preprint arXiv:1808.00046}, 2018.

\bibitem{cooke2006audio}
Martin Cooke, Jon Barker, Stuart Cunningham, and Xu~Shao.
\newblock An audio-visual corpus for speech perception and automatic speech
  recognition.
\newblock {\em The Journal of the Acoustical Society of America},
  120(5):2421--2424, 2006.

\bibitem{barker2015third}
Jon Barker, Ricard Marxer, Emmanuel Vincent, and Shinji Watanabe.
\newblock The third ‘chime’speech separation and recognition challenge:
  Dataset, task and baselines.
\newblock In {\em Automatic Speech Recognition and Understanding (ASRU), 2015
  IEEE Workshop on}, pages 504--511. IEEE, 2015.

\bibitem{viola2001rapid}
Paul Viola and Michael Jones.
\newblock Rapid object detection using a boosted cascade of simple features.
\newblock In {\em Computer Vision and Pattern Recognition, 2001. CVPR 2001.
  Proceedings of the 2001 IEEE Computer Society Conference on}, volume~1, pages
  I--I. IEEE, 2001.

\bibitem{ross2008incremental}
David~A Ross, Jongwoo Lim, Ruei-Sung Lin, and Ming-Hsuan Yang.
\newblock Incremental learning for robust visual tracking.
\newblock {\em International Journal of Computer Vision}, 77(1-3):125--141,
  2008.

\bibitem{ahmad2017compression}
Jawad Ahmad, Muazzam~A Khan, Seong~Oun Hwang, and Jan~Sher Khan.
\newblock A compression sensing and noise-tolerant image encryption scheme
  based on chaotic maps and orthogonal matrices.
\newblock {\em Neural Computing and Applications}, 28(1):953--967, 2017.

\bibitem{khan2017novel}
Fadia~Ali Khan, Jameel Ahmed, Jan~Sher Khan, Jawad Ahmad, and Muazzam~A Khan.
\newblock A novel substitution box for encryption based on lorenz equations.
\newblock In {\em Circuits, System and Simulation (ICCSS), 2017 International
  Conference on}, pages 32--36. IEEE, 2017.

\bibitem{khan2017td}
Jan~Sher Khan, Jawad Ahmad, and Muazzam~A Khan.
\newblock Td-ercs map-based confusion and diffusion of autocorrelated data.
\newblock {\em Nonlinear Dynamics}, 87(1):93--107, 2017.

\bibitem{ahmad2015chaos}
Jawad Ahmad and Seong~Oun Hwang.
\newblock Chaos-based diffusion for highly autocorrelated data in encryption
  algorithms.
\newblock {\em Nonlinear Dynamics}, 82(4):1839--1850, 2015.

\bibitem{anees2014chaotic}
Amir Anees, Adil~Masood Siddiqui, and Fawad Ahmed.
\newblock Chaotic substitution for highly autocorrelated data in encryption
  algorithm.
\newblock {\em Communications in Nonlinear Science and Numerical Simulation},
  19(9):3106--3118, 2014.

\bibitem{sathiyamurthi2017speech}
P~Sathiyamurthi and S~Ramakrishnan.
\newblock Speech encryption using chaotic shift keying for secured speech
  communication.
\newblock {\em EURASIP Journal on Audio, Speech, and Music Processing},
  2017(1):20, 2017.

\end{thebibliography}

\end{document}